\documentclass[aps,prb,twocolumn,superscriptaddress,nofootinbib,floatfix]{revtex4-2}

\usepackage{amsmath,amssymb,bm}
\usepackage{graphicx}
\usepackage{xcolor}
\usepackage[hidelinks]{hyperref}

\newcommand{\KGGp}{$K$-$\Gamma$-$\Gamma'$}

\newcommand{\kxy}{\kappa_{xy}}

\begin{document}

\title{Field-selected seven-site topological magnons in a classical frustrated triangular-lattice \texorpdfstring{$K$-$\Gamma$-$\Gamma'$}{K-Gamma-Gamma'} magnet}

\author{Bin Xi}
\email{xibin@yzu.edu.cn}
\affiliation{College of Physics Science and Technology, Yangzhou University, Yangzhou 225002, China}

\author{Jie Lu}
\affiliation{College of Physics Science and Technology, Yangzhou University, Yangzhou 225002, China}

\author{Shun-Li Yu}
\email{slyu@nju.edu.cn}
\affiliation{National Laboratory of Solid State Microstructures and School of Physics, Nanjing University, Nanjing 210093, China}
\affiliation{Collaborative Innovation Center of Advanced Microstructures, Nanjing University, Nanjing 210093, China}
\affiliation{Jiangsu Key Laboratory of Quantum Information Science and Technology, Nanjing University, Suzhou 215163, China}

\author{Yafang Xu}
\email{yfxu@yzu.edu.cn}
\affiliation{College of Physics Science and Technology, Yangzhou University, Yangzhou 225002, China}

\date{\today}

\begin{abstract}
Defining magnon topology in strongly frustrated magnets is often hindered by the absence of a simple harmonic magnon vacuum at zero field. Within a classical-spin ground-state search followed by linear spin-wave theory (LSWT), we demonstrate that a representative triangular-lattice $K$-$\Gamma$-$\Gamma'$ model in a dominant-$\Gamma$ exchange regime has an in-plane-field-selected compact, noncoplanar seven-site order. This field-selected state provides a controlled classical reference state and hosts magnon bands with field-tunable Chern numbers. Increasing the field drives a Dirac-like band touching that transfers Berry curvature between the fifth and sixth bands, altering the Chern vector from $(0,1,0,-2,1,0,0)$ to $(0,1,0,-2,-1,2,0)$. This topological transition reorganizes the band-resolved thermal Hall conductivity, driving the total $\kappa_{xy}(T)$ through a near-zero crossing once the upper bands are thermally populated. The dynamical structure factor places roughly half of the coherent spectral weight on the Chern-active branches, offering a spectroscopic route to identify the topological branches. These results define a controlled semiclassical benchmark for magnon topology in this pure nearest-neighbor exchange model.
\end{abstract}

\maketitle

\section{Introduction}

Magnon Chern numbers, chiral edge modes, and thermal Hall responses in linear
spin-wave theory (LSWT) are not properties of a spin Hamiltonian alone; they
require a stable harmonic expansion about a classically ordered, robust
magnetic vacuum whose magnetic unit cell defines the
Brillouin zone.  In frustrated magnets with competing
multi-$\mathbf{Q}$ states, the zero-field ground state often fails to provide
such a vacuum: independent Monte Carlo (MC) annealing runs can produce magnetic cells whose
sizes drift with the simulation cluster, so that band labels, direct gaps, and
Berry-curvature integrals depend on a finite-size commensurate
approximant---a cluster-size-dependent magnetic cell that approximates a
potentially incommensurate ordering---rather than on a bulk ordered phase.
The field-selection strategy used below
is to isolate from this frustrated zero-field manifold a robust reference
state with a small magnetic unit cell, after which magnon topology and
transport can be computed in a well-defined magnetic Brillouin zone.

Triangular-lattice spin-orbit magnets provide a natural setting for this
problem.  In these systems a spin-orbit-entangled local
doublet sits on three symmetry-related bonds, producing highly anisotropic, bond-dependent exchange
matrices with sizable Kitaev ($K$) or off-diagonal ($\Gamma$, $\Gamma'$)
terms~\cite{Li2015YbMgGaO4,Shen2016YbMgGaO4,Paddison2017YbMgGaO4,Ranjith2019NaYbO2,Kim2023CoI2,Xie2024CsCeSe2,Catuneanu2015Ba3IrTi2O9,Ortiz2023NaRuO2,Razpopov2023NaRuO2,Winter2017,RauGingras2018,Takagi2019,TrebstHickey2022,Rousochatzakis2024}.
The minimal nearest-neighbor \KGGp{} Hamiltonian contains two sources of
frustration: geometric frustration of the triangular lattice and
bond-dependent preferred spin components that cannot be optimized
simultaneously.  Their combination can generate multi-$\mathbf{Q}$ manifolds
and commensuration-sensitive ordered
states~\cite{Kitaev2006,JackeliKhaliullin2009,RauLeeKee2014,KimchiVishwanath2014,LiWangChen2016,LiuYuWang2016,LuoHuXi2017,ZhuMaksimov2018,KimchiNahumSenthil2018,ParkerBalents2018,JanssenVojta2019,WangQiXi2021,Maksimov2019,Rayyan2021,Chen2023TmX}.
In the $\Gamma$-dominant regime studied here, the zero-field states do not
converge to a fixed compact magnetic cell under a cluster-size scan
(Sec.~\ref{sec:zero_field}).

In existing studies of topological magnon bands, the ordered background is a
prerequisite for the topological calculation, where the central task is to diagonalize the spin-wave
Hamiltonian and evaluate its Berry
curvature~\cite{ZhangRenWangLi2013,Shindou2013,MookHenkMertig2014,MookEdge2014,Chisnell2015,Owerre2016Honeycomb,Pershoguba2018,McClartyDong2018,Joshi2018,McClarty2022,LiCaoYan2021}.
Within this framework, bond-dependent Kitaev-type interactions have been shown to generate magnon Chern numbers,
chiral edge modes, and thermal Hall sign structures in field-polarized and
other finite-cell phases of Kitaev
magnets~\cite{ZhangChernKim2021,ChernZhangKim2021,ChernCastelnovo2024}, and thermal Hall
transport provides a direct experimental probe of the underlying
time-reversal-odd
structure~\cite{Onose2010,Chisnell2015,Hirschberger2015Science,Hirschberger2015PRL,LaurellFiete2018,ChernBuessenKim2021}.
For the strongly frustrated \KGGp{} regime, the ordered background must first
be selected and tested before topological band quantities are assigned.

We address this issue in the triangular-lattice \KGGp{} model, using it as a
minimal setting motivated by the triangular-lattice Kitaev candidate NaRuO$_2$~\cite{Ortiz2023NaRuO2,Razpopov2023NaRuO2,Bhattacharyya2023NaRuO2}.  We
focus on the representative $\Gamma$-dominant parameter point fixed in
Sec.~\ref{sec:model_methods}.  At this parameter point, the zero-field states do not
provide a robust spin-wave vacuum.  An in-plane field along the
crystallographic $a$ axis instead selects a compact noncoplanar seven-site
order over a finite field window.  We test this selection using
multi-seed ground-state searches, an $H\parallel c$ control, an in-plane
angular scan, a local exchange-parameter scan, and compatible-torus checks.

Within this field-selected state, the seven magnon bands carry nontrivial
Chern numbers and undergo a field-tunable topological transition: a
Dirac-like contact between bands 5 and 6 transfers Berry curvature and alters
the Chern vector.  The same transfer reorganizes the band-resolved thermal
Hall conductivity, driving the total $\kxy(T)$ through a near-zero crossing
once the upper bands are thermally populated.  The dynamical structure factor
computed within LSWT shows that roughly half of the coherent spectral weight
averaged along the path resides on the Chern-active branches, offering a
spectral route to the band assignment.

The paper is organized as follows.  Section~\ref{sec:model_methods} fixes the
model, coordinate conventions, and numerical machinery.
Section~\ref{sec:zero_field} documents why the zero-field states are not used
as topological LSWT reference states.  Section~\ref{sec:field_selection}
establishes the field-selected seven-site state, its angular selection window,
and its robustness under local parameter scans and compatible-torus checks.
Section~\ref{sec:magnons} reports the magnon bands, Chern numbers, edge modes,
and thermal Hall response, and Sec.~\ref{sec:dynamics} the dynamical structure
factor.  Section~\ref{sec:discussion} discusses implications and limitations.

\section{Model and methods}
\label{sec:model_methods}

\subsection{Triangular-lattice \texorpdfstring{$K$-$\Gamma$-$\Gamma'$}{K-Gamma-Gamma'} model}

We consider classical spins of fixed length $|\mathbf{S}_i|=1$ on a triangular
lattice.  Nearest-neighbor bonds are divided into three symmetry-related types,
denoted by $\gamma=x,y,z$.  In the spin component basis used throughout the
calculation, the Hamiltonian is
\begin{equation}
  \mathcal{H}
  =
  \sum_{\langle ij\rangle_\gamma}
  \mathbf{S}_i^{\mathsf T} M_\gamma \mathbf{S}_j
  - \sum_i \mathbf{H}\cdot\mathbf{S}_i ,
  \label{eq:hamiltonian}
\end{equation}
where
\begin{align}
M_x &=
\begin{pmatrix}
K & \Gamma' & \Gamma' \\
\Gamma' & 0 & \Gamma \\
\Gamma' & \Gamma & 0
\end{pmatrix}, \nonumber\\
M_y &=
\begin{pmatrix}
0 & \Gamma' & \Gamma \\
\Gamma' & K & \Gamma' \\
\Gamma & \Gamma' & 0
\end{pmatrix}, \nonumber\\
M_z &=
\begin{pmatrix}
0 & \Gamma & \Gamma' \\
\Gamma & 0 & \Gamma' \\
\Gamma' & \Gamma' & K
\end{pmatrix}.
\label{eq:bond_matrices}
\end{align}
The sum runs over undirected nearest-neighbor bonds, each counted once.  We
parameterize the anisotropic exchanges as
\begin{equation}
  K=\sin\phi\sin\theta,\qquad
  \Gamma=\cos\phi\sin\theta,\qquad
  \Gamma'=\cos\theta ,
  \label{eq:parametrization}
\end{equation}
so that $K^2+\Gamma^2+\Gamma'^2=1$ by construction.  All coupling triples quoted
in this manuscript follow Eq.~\eqref{eq:parametrization}.  The field-selected
state is studied at $(\theta,\phi)=(0.55\pi,1.91\pi)$, for which
\begin{equation}
  (K,\Gamma,\Gamma')=(-0.27556,\;0.94847,\;-0.15643).
\end{equation}
Equation~\eqref{eq:hamiltonian} is dimensionless. For a quantum model with
exchange scale $E_0$ and spin $S$ the dimensionless field is related to the
physical field $B$ by
\begin{equation}
  H=\frac{g\mu_B B}{E_0S}.
  \label{eq:field_units}
\end{equation}

The field direction is specified in the crystallographic $(a,b,c)$ frame,
whereas Eq.~\eqref{eq:bond_matrices} is written in the cubic spin basis
$(x,y,z)$.  We use
\begin{align}
  \hat{\mathbf a} &= (1,1,-2)/\sqrt{6}, \nonumber\\
  \hat{\mathbf b} &= (-1,1,0)/\sqrt{2}, \nonumber\\
  \hat{\mathbf c} &= (1,1,1)/\sqrt{3},
  \label{eq:abc_axes}
\end{align}
which form a right-handed orthonormal triad.  An in-plane field angle $\alpha$
therefore denotes
\begin{equation}
  \mathbf{H}
  =
  H\left(\cos\alpha\,\hat{\mathbf a}
  +\sin\alpha\,\hat{\mathbf b}\right).
  \label{eq:inplane_field}
\end{equation}
Thus $\alpha=0^\circ$ corresponds to the crystallographic $a$ direction,
while $\alpha=90^\circ$ corresponds to the $b$ direction.  This convention is
kept explicit in all field-angle scans.

The spin-wave and transport calculations below use the leading harmonic term
in the large-$S$ expansion; possible corrections beyond LSWT are discussed in
Sec.~\ref{sec:discussion}.

\subsection{Parallel-tempering annealing and quenching}

Ground-state searches are performed by a finite-temperature
parallel-tempering stage followed by a zero-temperature local-field
quench~\cite{HukushimaNemoto1996}.  Each MPI rank carries a replica at one
temperature in a grid between $T_{\min}$ and $T_{\max}$.  Neighboring
temperature labels are exchanged with the Metropolis probability determined by
their energy and inverse-temperature differences.  During the subsequent quench
stage, each spin is iteratively aligned with its effective field
$\mathbf{h}^{\rm eff}_i$, defined by the single-spin energy
$E_i=-\mathbf{h}^{\rm eff}_i\cdot\mathbf{S}_i$, until a locally
self-consistent configuration is reached.

The main field-window scan for $H\parallel a$ uses clusters
$L=6,8,\ldots,42$ and three independent random seeds, with temperatures on
a quadratic grid between $T_{\min}=0.002$ and $T_{\max}=1.2$ and $10^5$
Monte Carlo updates in each of the parallel-tempering and quench stages.
A magnetic cell is retained only when independent seeds reproduce the same
energy and the same cell, up to symmetry-equivalent choices of
magnetic primitive vectors.  An $H\parallel c$ comparison uses the same
cluster range and seed criterion at
$H=0.0500,0.1000,0.1500,0.2000$ and
$H=0.2200,0.2500,0.2625,0.3000$.

\subsection{Magnetic-cell detection}

For each relaxed spin configuration we identify primitive magnetic
translations
$\mathbf{T}_1=m_1\mathbf{a}_1+n_1\mathbf{a}_2$,
$\mathbf{T}_2=m_2\mathbf{a}_1+n_2\mathbf{a}_2$ that reproduce the spin
pattern up to a numerical tolerance, with magnetic cell size
$N_{\rm mag}=|m_1 n_2-m_2 n_1|$.  The corresponding magnetic Brillouin
zone is constructed from the dual reciprocal basis.  Band and Berry-curvature
calculations are carried out in the magnetic zone, while the dynamical
structure factor is plotted along the physical triangular-lattice
$\Gamma$-$M$-$K$-$\Gamma$ path and folded into the magnetic zone for the
LSWT eigenproblem.  We use ``$k$-site'' as shorthand for a genuine
two-dimensional magnetic cell with $N_{\rm mag}=k$; in particular, a
one-site entry is a uniform ferromagnetic (FM) state with the primitive
crystallographic periodicity.  This is distinct from a rank-one pattern,
where one magnetic translation is detected but no independent second
translation is found within the search window.  Algorithmic details and the explicit reciprocal-vector
convention are given in Appendix~\ref{app:cell_detection}.

\subsection{Linear spin-wave theory}

For each locally stable classical state we set up a local orthonormal frame
with $\hat z$ along the classical spin direction at every basis site.
Holstein-Primakoff expansion to quadratic order yields a Bogoliubov-de Gennes
bosonic Hamiltonian $\mathcal{H}_2=\tfrac{1}{2}\sum_{\mathbf k}
\Psi_{\mathbf k}^\dagger \mathcal{M}(\mathbf k)\Psi_{\mathbf k}$ in the
Nambu basis
$\Psi_{\mathbf k}=(a_{1\mathbf k},\ldots,a_{N\mathbf k},
a^\dagger_{1,-\mathbf k},\ldots,a^\dagger_{N,-\mathbf k})^{\mathsf T}$,
where $N=N_{\rm mag}$ is the number of basis sites in the magnetic cell.
The associated bosonic paraunitary metric is
$\eta=\mathrm{diag}(I_N,-I_N)$, with $I_N$ the $N\times N$ identity matrix.
We diagonalize this problem by the standard Colpa
procedure~\cite{Colpa1978,TothLake2015}.  We use a magnetic-cell Bloch gauge in
which the bond translation carries the only
momentum-dependent phase, so that
$\mathcal{M}(\mathbf{k}+\mathbf{G}_{\rm mag})=\mathcal{M}(\mathbf{k})$ to
numerical precision, as required for a periodic Chern-number calculation.  We
accept a state as a harmonic magnon vacuum only when the eigenvalue imaginary parts
are at machine precision and the minimum magnon frequency is strictly
positive.

\subsection{Chern numbers and thermal Hall response}

Band Chern numbers $C_n$ are computed by the Fukui-Hatsugai-Suzuki
discretization~\cite{Fukui2005} on a uniform mesh in the magnetic Brillouin
zone.  For bosonic systems the lattice Berry connection is built from the
paraunitary overlap
$U_n(\mathbf{k},\mathbf{k}+\boldsymbol{\delta})=[T_{\mathbf{k}}^\dagger\eta\,
T_{\mathbf{k}+\boldsymbol{\delta}}]_{nn}$ (the Berry-link variable), where
$T_{\mathbf{k}}$ is the paraunitary matrix diagonalizing the Bogoliubov
Hamiltonian and $\eta$ is the bosonic metric.  We
quote band-resolved Chern numbers only when adjacent direct gaps remain open on
the momentum mesh and report the minimum Berry-link modulus as a singularity
diagnostic.

The dimensionless magnon thermal Hall conductivity is evaluated in the
Matsumoto-Murakami
convention~\cite{Katsura2010,QinNiuShi2011,MatsumotoMurakami2011PRL,MatsumotoMurakami2011,MurakamiOkamoto2017,Shindou2013}:
\begin{equation}
  \kxy
  =
  -\frac{T}{(2\pi)^2}
  \sum_{n,\mathbf{k}}
  \left[
    c_2(\rho_{n\mathbf{k}})-\frac{\pi^2}{3}
  \right]
  F_n(\mathbf{k}),
  \label{eq:kappa}
\end{equation}
where $\rho_{n\mathbf k}=(e^{\omega_{n\mathbf k}/T}-1)^{-1}$, $c_2$ is the
weight function of Ref.~\onlinecite{MatsumotoMurakami2011PRL}, and
$F_n(\mathbf k)$ is the lattice Berry flux per plaquette of the
$N_k\times N_k$ mesh in reduced magnetic reciprocal coordinates.  We use
$k_B=\hbar=1$ throughout.

\subsection{Dynamical structure factor}

For the seven-site state we compute the coherent one-magnon dynamical
structure factor from the LSWT eigenmodes,
\begin{equation}
  S(\mathbf{q},\omega)
  =
  \sum_n I_n(\mathbf{q})\,
  \delta_\eta\!\left(\omega-\omega_{n\mathbf{q}}\right),
  \label{eq:lswt_sqw}
\end{equation}
with $I_n(\mathbf{q})=\sum_{\alpha\beta}(\delta_{\alpha\beta}-\hat{q}_\alpha
\hat{q}_\beta)\langle 0|S^{\alpha}_{\mathbf{q}}|n\mathbf{q}\rangle\langle
n\mathbf{q}|S^{\beta}_{-\mathbf{q}}|0\rangle$ the polarization-averaged
spin-flip matrix element in the global $(x,y,z)$ basis.  Plots use a
Gaussian broadening $\delta_\sigma(\omega-\omega_{n\mathbf{q}})=
(\sigma\sqrt{2\pi})^{-1}e^{-(\omega-\omega_{n\mathbf{q}})^2/2\sigma^2}$
with $\sigma=0.04$ (in dimensionless frequency units) for visualization;
the band weights quoted in Sec.~\ref{sec:dynamics} are computed from the
unbroadened mode intensities $I_n(\mathbf{q})$.

\section{Zero-field obstruction to a compact magnetic cell}
\label{sec:zero_field}

\begin{figure}[!htbp]
  \centering
  \includegraphics[width=\columnwidth]{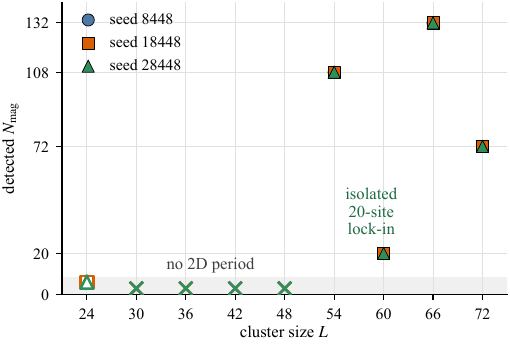}
  \caption{
  Zero-field magnetic-cell detection results at
  $(\theta,\phi)=(0.55\pi,1.91\pi)$.  Detected magnetic-cell size versus
  cluster size for three independent seeds.  Filled markers denote detected
  two-dimensional finite cells, open markers denote rank-one patterns, and
  crosses mark clusters where no two-dimensional period is found within the
  search window.  The isolated 20-site cell at $L=60$ is a finite-size
  commensurate lock-in.
  }
  \label{fig:zero_field}
\end{figure}

At $(\theta,\phi)=(0.55\pi,1.91\pi)$, the zero-field state does not
provide a robust harmonic magnon vacuum.  Independent annealing and
quench runs on clusters with $L=24$--$72$ do not produce a sequence of
magnetic cells converging to a fixed magnetic unit cell.  Small clusters
either give a rank-one pattern or no two-dimensional translation within the
search window, while the finite cells detected at larger sizes jump between
$N_{\rm mag}=108$, 20, 132, and 72.  The isolated 20-site cell at $L=60$ is
therefore a finite-size
commensurate lock-in rather than evidence for a genuine bulk phase
[Fig.~\ref{fig:zero_field}; Appendix~\ref{app:zero_field_cells}].
This finite-size scan leaves no unique periodic magnetic order on which to
formulate the LSWT eigenvalue problem.
The LSWT stability diagnostics sharpen this conclusion.  The $L=54$
approximants, which produce a 108-site detected cell, have complex spin-wave
eigenvalues with maximum imaginary part $\simeq4.9\times10^{-2}$ on a
$21\times21$ magnetic-zone mesh.  The isolated 20-site lock-in at $L=60$ is
harmonically stable with $\omega_{\min}=0.190$, while the 132- and 72-site
approximants at $L=66$ and $72$ are positive but very soft
($\omega_{\min}\sim10^{-3}$).  Thus the positive spectra that do occur belong
to mutually incompatible finite-size lock-ins rather than to a converged bulk
ordered phase, and we do not assign zero-field magnon Chern numbers.

Because the approximants are mutually incompatible and do not converge under
the cluster-size scan, they are not used as bulk orders for topology.
The field-selected compact order is described next.

\section{Field-selected magnetic order}
\label{sec:field_selection}

\begin{figure}[!htbp]
  \centering
  \includegraphics[width=\columnwidth]{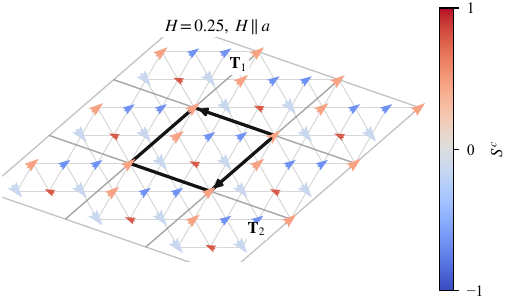}
  \caption{
  The field-selected state is a compact noncoplanar seven-site order.
  Real-space spin texture at $H=0.25$ for $H\parallel a$, obtained by
  periodically repeating the magnetic cell used in the spin-wave calculation.
  The black parallelogram marks one magnetic unit cell, with
  $\mathbf{T}_1=(-3,1)$ and $\mathbf{T}_2=(-1,-2)$ in triangular-lattice
  coordinates.  Arrows show $(S^a,S^b)$ in the crystallographic frame and the
  color gives $S^c$.
  }
  \label{fig:ha7_cell}
\end{figure}

The field direction determines whether a compact magnetic cell is stabilized.
An in-plane $a$-type field near $H=0.25$
selects a compact seven-site noncoplanar order, while the high-symmetry
$c$-axis field ($\hat{\mathbf{c}}\parallel(1,1,1)$ in the cubic spin basis)
and generic in-plane directions do not.  One convenient set of magnetic
primitive translations for this state is
\begin{align}
  \mathbf{T}_1 &= -3\mathbf{a}_1+\mathbf{a}_2,\nonumber\\
  \mathbf{T}_2 &= -\mathbf{a}_1-2\mathbf{a}_2,
  \label{eq:seven_site_cell}
\end{align}
with $|\det(\mathbf{T}_1,\mathbf{T}_2)|=7$.  The corresponding spin texture
is shown in Fig.~\ref{fig:ha7_cell}: the order is noncollinear and
noncoplanar, and independent seeds reproduce it up to symmetry-equivalent
domains.

We establish the field window and direction selectivity using parallel
tempering followed by zero-temperature local-field quenching on clusters
$L=6,8,\ldots,42$ with three independent random seeds; a candidate ground
state is accepted only when all three seeds reproduce the same energy and
magnetic unit cell, allowing symmetry-equivalent primitive translations.
For $H\parallel a$ this procedure yields the sequence
\begin{equation}
  20\text{-site}
  \;\rightarrow\;
  7\text{-site}
  \;\rightarrow\;
  36\text{-site},
  \label{eq:field_cell_sequence}
\end{equation}
with the seven-site cell stable for $0.2200\le H\le 0.2625$.  The lower
boundary lies between $H=0.2150$ and $0.2200$ (seeds split, 20-site slightly
lower); the upper boundary lies between $0.2625$ and $0.2650$, where
$\Delta\varepsilon=\varepsilon_{36}-\varepsilon_{7}$ reverses sign and the
field-parallel magnetization jumps from $m_\parallel=0.495$ to $0.577$,
signaling a first-order transition to the 36-site branch.  A control scan for
$H\parallel c$ from $H=0.0500$ to $0.3000$ remains on the 20-site branch
across all three seeds, indicating that the seven-site stability window is tied to the
field direction rather than the field magnitude.  A fixed-field in-plane angle
scan gives a complementary check: at $H=0.25$, a $5^\circ$ deviation from the
$a$ axis already selects the 36-site branch, and the seven-site state appears
only near $\alpha=0^\circ$ and the symmetry-related $\alpha=60^\circ$
direction (Table~S1 of the Supplemental Material~\cite{SupplementalMaterial}).
The angular selection follows from the seven-site translation lattice.  The
translations in Eq.~\eqref{eq:seven_site_cell} are closed under proper
triangular-lattice rotations: a $60^\circ$ rotation maps $\mathbf{T}_1$ to
$\mathbf{T}_2$ and $\mathbf{T}_2$ to $-\mathbf{T}_1+\mathbf{T}_2$.  The two
$a$-type directions $\alpha=0^\circ$ and $60^\circ$ therefore share the same
index-seven translation lattice, while mirror operations produce the opposite
chiral domain.  Generic in-plane field angles lack this closure and select
competing commensurate cells.

\begin{figure}[!htbp]
  \centering
  \includegraphics[width=\columnwidth]{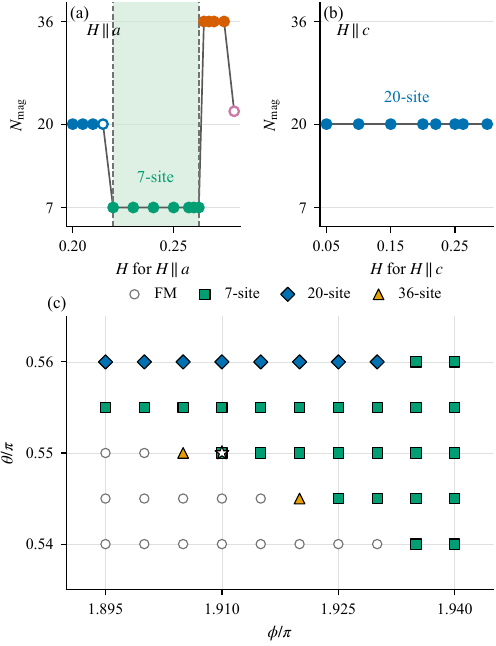}
  \caption{
  Field and parameter diagnostics for the selected seven-site state.
  (a) Magnetic-cell size selected for $H\parallel a$.  Filled (open) markers
  denote fields where independent seeds agree (split), and the shaded band
  marks $0.2200\le H\le0.2625$.
  (b) Control scan for $H\parallel c$: fields from $0.0500$ to $0.3000$ remain
  on a 20-site branch across all three seeds.
  (c) Single-seed local scan at $H=0.25$ and $H\parallel a$.  Symbols mark the
  selected two-dimensional magnetic cell at each sampled parameter point; FM
  denotes the one-site uniform ferromagnetic branch and the star marks
  $(\theta,\phi)=(0.55\pi,1.91\pi)$.
  }
  \label{fig:field_selection}
\end{figure}

A local exchange-parameter scan places the working point within a finite
seven-site region.  At fixed $H=0.25$ and $H\parallel a$, a single-seed grid in
$0.540\le\theta/\pi\le0.560$ and $1.895\le\phi/\pi\le1.940$ finds the
seven-site cell in 25 of the 50 sampled points [Fig.~\ref{fig:field_selection}(c)].
The working point lies inside a connected seven-site region on this grid:
lower-$\theta$ and lower-$\phi$ points tend toward the FM one-site
branch, the upper side is adjacent to a 20-site branch, and two boundary
points select a 36-site cell.  This finite-cluster scan is local in
parameter space and does not constitute a thermodynamic phase diagram.

Representative LSWT spectra of the neighboring 20-site and 36-site branches
are shown in Fig.~S2 of the Supplemental Material~\cite{SupplementalMaterial}.  They
are locally harmonically stable, but
their dense band manifolds contain adjacent-band direct gaps of order
$10^{-4}$--$10^{-3}$.  In the present work these spectra serve to characterize
the neighboring field-selected branches, while the topology and transport
analysis below focuses on the seven-site state.

A finite-size compatibility check is consistent with the seven-site branch.
Compatible tori ($L=14,28,42$) reproduce the same energy per site,
$E_7/N=-1.17905$, to roundoff, while incompatible sizes lie higher because
they cannot tile the seven-site cell exactly (Appendix~\ref{app:zero_field_cells};
Fig.~S1 of the Supplemental Material~\cite{SupplementalMaterial}).  The
compatible sizes show no detectable drift with $1/L^2$.

The noncoplanarity of the seven-site state is quantified by the scalar
chirality on nearest-neighbor triangles,
$\chi_{ijk}=\mathbf{S}_i\cdot(\mathbf{S}_j\times\mathbf{S}_k)$,
with sites ordered counterclockwise around each elementary triangle in the
crystallographic $ab$ plane.  Averaging over the fourteen elementary triangles
per magnetic cell gives $\langle\chi\rangle=-0.260$,
$\langle|\chi|\rangle=0.351$, and $\sqrt{\langle\chi^2\rangle}=0.497$, with
maximum $|\chi|=0.914$.  This sizable chirality is consistent with
the nontrivial magnon Berry curvature analyzed in Sec.~\ref{sec:magnons};
reversing the field together with all spins reverses the scalar chirality and
the associated time-reversal-odd transport response.

\section{Topological magnons and thermal Hall response}
\label{sec:magnons}

We use the field-selected $H\parallel a$ seven-site branch as the classical
reference state and compute its magnon bands, Berry curvature, and thermal Hall
response.  The seven nondegenerate bands of the
magnetic supercell each carry an integer Chern number constrained by
$\sum_n C_n=0$~\cite{Shindou2013}.  Over the field range $0.2500\le H\le
0.2625$ used throughout this section, the LSWT spectrum remains positive throughout the full magnetic Brillouin
zone: the minimum magnon energy falls only from $0.3438$
to $0.3220$, the bosonic eigenvalue imaginary parts remain at machine
precision, and the minimum Berry-link modulus on the $121\times121$ mesh
remains far from zero: it reaches a minimum of $0.32$ at
$H=0.2575$ (adjacent to the Chern-transfer interval) and rises to $0.81$ at
$H=0.2625$.  Throughout this range, the Chern-number calculation remains well
conditioned and the seven-site state
remains a positive-energy harmonic magnon vacuum throughout this field window.

\subsection{Magnon bands and Chern transfer}

\begin{figure}[!htbp]
  \centering
  \includegraphics[width=\columnwidth]{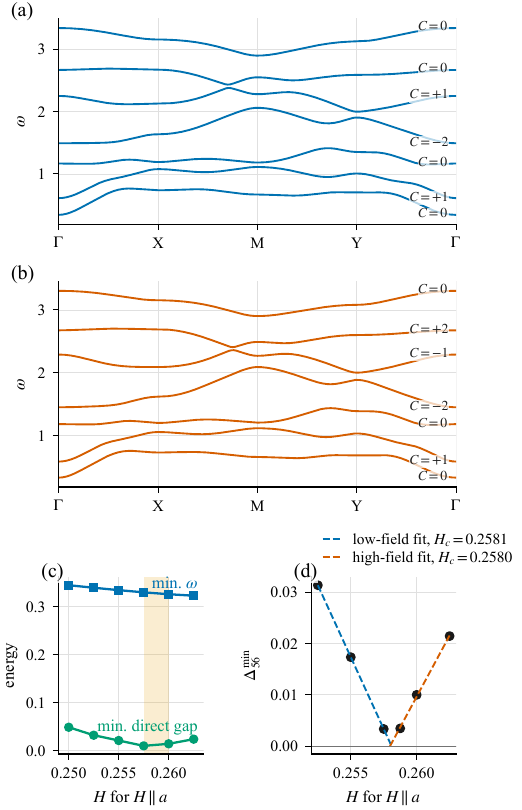}
  \caption{
  The selected seven-site state carries field-tunable magnon Chern numbers.
  (a) Linear-spin-wave spectrum at $H=0.2500$, with Chern vector
  $\mathbf{C}_{\rm low}$, and (b) spectrum at $H=0.2600$, with Chern vector
  $\mathbf{C}_{\rm high}$.  Small labels at the right edge mark the band Chern
  numbers.  Both spectra are plotted along the magnetic
  Brillouin-zone path $\Gamma$-$X$-$M$-$Y$-$\Gamma$.  This path is defined in
  the reduced coordinates of the seven-site magnetic Brillouin zone, with
  $X=(1/2,0)$, $M=(1/2,1/2)$, and $Y=(0,1/2)$ in the reciprocal basis of the
  magnetic supercell.  In physical reciprocal coordinates these points are
  $X=(-1/7,1/14)$, $M=(-3/14,-1/7)$, and $Y=(-1/14,-3/14)$ modulo reciprocal
  lattice vectors, distinct from the physical $M$ and $K$ points used below.
  (c) The minimum positive magnon frequency remains finite while the smallest
  adjacent-band direct gap closes near the transfer.  The shaded interval
  marks the field range in which the fifth and sixth bands exchange Chern
  number.  (d) Minimum direct gap between bands 5 and 6 obtained from the
  adaptive momentum search, with linear fits on the two sides.  Fields above
  $H=0.2625$ are not included in the seven-site branch because
  the independent ground-state search selects larger magnetic cells there.
  }
  \label{fig:topological_magnons}
\end{figure}

Figure~\ref{fig:topological_magnons} shows the magnon bands and Chern
numbers on the seven-site branch.  At $H=0.2500$ the Chern vector is
\begin{equation}
  \mathbf{C}_{\rm low}
  =
  (0,1,0,-2,1,0,0),
  \label{eq:chern_low_field}
\end{equation}
and between $H=0.2575$ and $H=0.2600$ it becomes
\begin{equation}
  \mathbf{C}_{\rm high}
  =
  (0,1,0,-2,-1,2,0),
  \label{eq:chern_high_field}
\end{equation}
through a Chern transfer between bands 5 and 6 in which $C_5$ flips from
$+1$ to $-1$ and $C_6$ jumps from $0$ to $+2$.  Both vectors satisfy
$\sum_n C_n=0$.  The remaining bands keep their Chern numbers across the
transfer; in particular the band-2 ($C=+1$) and band-4 ($C=-2$) sectors
persist on both sides of the window.

The transfer occurs through Dirac-like contacts at generic momenta.  The
band-5/band-6 direct gap closes linearly in $|H-H_c|$ from both sides:
linear fits to the lower- and upper-field points give
$H_c\simeq 0.2581$ and $H_c\simeq 0.2580$, respectively.  These estimates
place the transition at $H_c\simeq 0.258$ to within the field-grid resolution
[Fig.~\ref{fig:topological_magnons}(d)].  The gap minimum lies near
$(k_1,k_2)\approx(0.46,0.37)$ up to $H=0.25875$ and shifts to
$(0.55,0.63)$ by $H=0.2600$, expressed in the fractional coordinates of the
magnetic Brillouin zone.  The gap does not close at the $M$ point of the
$\Gamma$-$X$-$M$-$Y$-$\Gamma$ path, consistent
with a generic rather than high-symmetry band touching.
The net change $\Delta C_5=-2$ and $\Delta C_6=+2$ is consistent with two
generic Dirac-like contacts, approximately related by
$(k_1,k_2)\mapsto(1-k_1,1-k_2)$ modulo magnetic reciprocal lattice vectors,
each exchanging Berry flux $\pm\pi$ between the two bands and contributing
$\Delta C=\mp1$~\cite{Haldane1988}.
The integer Chern assignments are stable on finer momentum meshes at
$H=0.2500,0.2575,0.2600$, and $0.2625$.

\subsection{Edge states}

\begin{figure}[!htbp]
  \centering
  \includegraphics[width=\columnwidth]{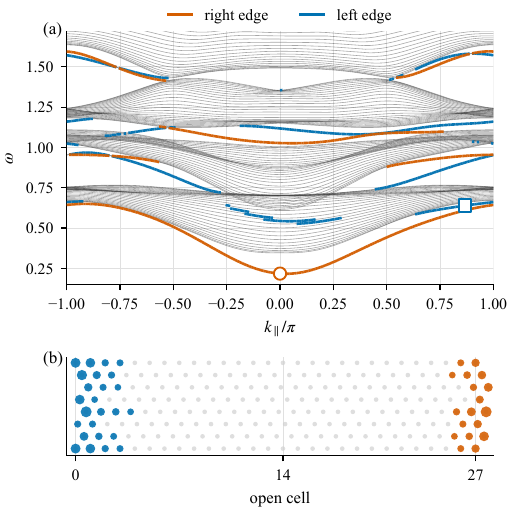}
  \caption{
  Low-energy strip spectrum at $H=0.2500$.  (a) Strip spectrum versus
  momentum along the periodic direction.  Thin grey lines show strip bands in
  the plotted energy window; red and blue segments mark states whose weight on
  the two outermost magnetic cells exceeds $0.35$ on opposite boundaries.  Open
  symbols mark the two modes used for the real-space map.  (b) Site-resolved
  magnon weights for representative left- and right-edge modes on the same
  strip.
  }
  \label{fig:edge_states}
\end{figure}

We use a seven-site strip to visualize the boundary response associated
with the low-field Chern vector~\cite{Hatsugai1993,Shindou2013}.  The strip is periodic along
$\mathbf{T}_1=(-3,1)$ and open across $28$ magnetic cells along
$\mathbf{T}_2=(-1,-2)$, with hard-wall boundaries implemented by omitting
bonds that leave the strip.  The bosonic spectrum remains positive
($\omega_{\min}=0.216$, eigenvalue imaginary parts at machine precision).
The periodic bulk calculation gives
$\mathbf{C}_{\rm low}=(0,1,0,-2,1,0,0)$, producing nonzero cumulative Chern
numbers in the low-energy gaps.  In the strip spectrum these boundary
modes appear as edge-weighted branches between the projected bulk bands
(Fig.~\ref{fig:edge_states}).

The highlighted branches are localized on opposite edges and disperse with
opposite group velocities.  The two representative modes marked in the
spectrum have edge weights $0.955$ and $0.921$ on the two outermost magnetic
cells, and their real-space weights confirm localization at opposite
boundaries.  The colored branches are therefore boundary-localized strip
modes associated with the nonzero bulk Chern sums.  Surface spins are not
relaxed in this construction, so boundary relaxation can shift edge-mode
energies without changing the bulk Chern numbers.

\subsection{Thermal Hall response}

The thermal Hall response probes the Berry curvature
within the seven-site LSWT model.  Since Eq.~\eqref{eq:kappa} weights each
band by the Bose occupation, it is not determined by the Chern vector alone.
The low-energy bands dominate first, while the higher bands that participate
in the Chern transfer enter only as the temperature increases.  This makes
$\kxy$ sensitive both to the Berry-curvature sign and to the energy scale of
the bands that carry that curvature.

\begin{figure}[!htbp]
  \centering
  \includegraphics[width=\columnwidth]{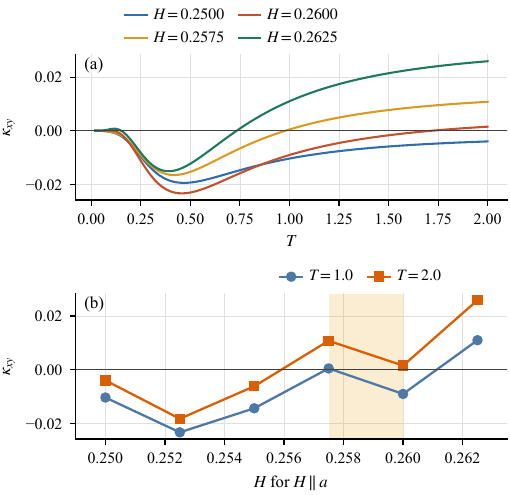}
  \caption{
  The total thermal Hall response passes through a near-zero value in the
  Chern-transfer field range once the transfer-active bands are thermally
  populated.
  (a) Temperature dependence of the dimensionless $\kxy$ for representative
  fields across the seven-site window.
  (b) Field dependence at fixed temperatures $T=1.0$ and $T=2.0$.  The shaded
  interval is the same Chern-transfer region identified in
  Fig.~\ref{fig:topological_magnons}.  The near-zero crossing at high
  temperature occurs close to the lower edge of this interval, while the
  topological transition further redistributes thermally weighted Berry
  curvature between bands 5 and 6.
  Temperature is dimensionless (units of $E_0 S/k_B$); for
  $E_0=5$--$10$\,meV and $S=1/2$ the unit $T=1$ corresponds to
  $T_{\rm phys}\approx29$--$58$\,K.
  }
  \label{fig:thermal_hall}
\end{figure}

The band decomposition shows that the total response is controlled by a
near-cancellation among a few Chern-active bands.  Below the transfer, at
$H=0.2575$ and $T=2$, the dominant terms are
$\kxy^{(4)}=-0.540$, $\kxy^{(2)}=+0.168$, and
$\kxy^{(5)}=+0.365$, with a small $\kxy^{(6)}=+0.020$; their sum is only
$\kxy=+0.0108$.  Above the transfer, at $H=0.2625$, bands 2 and 4 change
little, while the Chern-active upper pair is reorganized:
$\kxy^{(5)}=-0.354$ and $\kxy^{(6)}=+0.743$, giving
$\kxy=+0.0259$.  The full band-resolved $\kxy^{(n)}(T)$ curves are shown
in Fig.~S3 of the Supplemental Material~\cite{SupplementalMaterial}.

This cancellation accounts for the temperature and field dependences in
Fig.~\ref{fig:thermal_hall}.  At low temperature, the upper bands are
thermally suppressed and the total response remains negative and weakly
field-dependent across the Chern-transfer interval.  At larger temperature,
the band-5/band-6 reorganization enters the transport sum and shifts the
total response through a near-zero crossing between $H=0.2550$ and
$0.2575$.  The thermal Hall curve therefore provides a transport signature of
the Chern transfer through thermally weighted Berry curvature rather than
through the integer Chern labels alone.

\section{Dynamical structure factor}
\label{sec:dynamics}

The dynamical structure factor (DSF) resolves the Chern-active branches in
energy and momentum, providing a spectroscopic complement to the
Berry-curvature analysis of the previous section.

\begin{figure}[!htbp]
  \centering
  \includegraphics[width=\columnwidth]{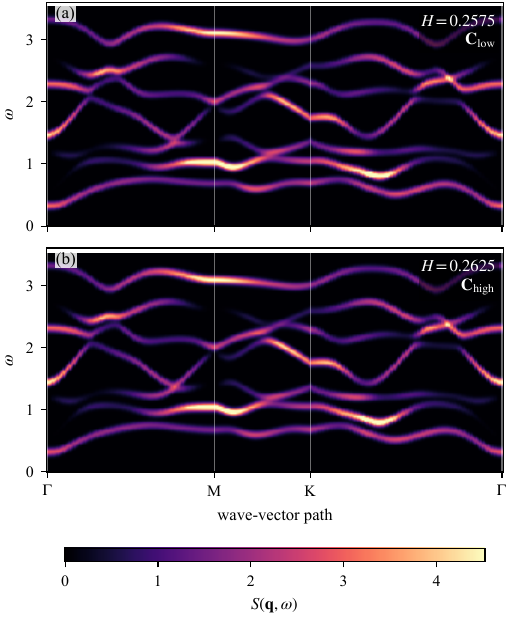}
  \caption{
  The dynamical structure factor tracks the magnon branches across the Chern
  transfer while retaining band-selective spectral weight.
  LSWT $S(\mathbf{q},\omega)$ on the two sides of the Chern-transfer interval:
  (a) $H=0.2575$, with Chern vector
  $(0,1,0,-2,1,0,0)$, and (b) $H=0.2625$, with Chern vector
  $(0,1,0,-2,-1,2,0)$.  Both panels use the same color scale and are plotted
  along the physical triangular-lattice $\Gamma$-$M$-$K$-$\Gamma$ path.
  Physical wave vectors are folded into the magnetic Brillouin zone for the
  LSWT eigenproblem, while the intensity carries the physical-momentum phase factor.
  }
  \label{fig:dynamical_structure_factor}
\end{figure}

Figure~\ref{fig:dynamical_structure_factor} plots the coherent LSWT response
$S(\mathbf{q},\omega)$ immediately before ($H=0.2575$) and after
($H=0.2625$) the Chern-transfer interval.  The spectra follow the standard
$\Gamma$-$M$-$K$-$\Gamma$ path of the physical triangular lattice, with
physical wave vectors folded into the magnetic Brillouin zone for the
eigenproblem, while the intensity carries the physical-momentum phase factor.  The
magnon branches remain sharp and change continuously across the transfer,
indicating that the Chern relabeling between bands 5 and 6 is not accompanied
by a large redistribution of coherent spectral weight.

The component-resolved spectra in Fig.~S4 of the Supplemental
Material~\cite{SupplementalMaterial} show the same branch structure in the
crystallographic $a$, $b$, and $c$ spin channels.  Thus the visible
Chern-active branches are not selected by a single polarization channel,
although their relative intensities are component dependent.

The unbroadened band weights at the central window point $H=0.2500$ provide a
quantitative check of this identification.  As summarized in
Table~\ref{tab:dsf_weights}, the low-field Chern-active bands 2, 4, and 5
carry $\simeq 47.5\%$ of the coherent weight averaged along the path and nearly
$60\%$ near $\Gamma$ and $M$.  The transfer-active bands 5 and 6 lie higher
in the spectrum, at $\omega\sim2.3$--$2.6$; band 5 remains visible with
$\sim14\%$ weight averaged along the path, whereas band 6 is weaker, with
$\sim9\%$ and little weight at $\Gamma$ and $M$.  Thus the DSF identifies
the visible Chern-active branches, while the thermal Hall response remains
a more direct probe of the Berry-curvature transfer itself.

\begin{table}[!htbp]
  \centering
  \caption{
  Unbroadened LSWT spectral-weight fractions $I_n(\mathbf{q})/\sum_n I_n$
  along the physical $\Gamma$-$M$-$K$-$\Gamma$ path of the $H=0.2500$
  seven-site state.  ``avg'' denotes the fraction averaged along the path.
  Chern labels
  refer to $\mathbf{C}_{\rm low}$ in Eq.~\eqref{eq:chern_low_field}.
  Topologically nontrivial bands are marked with bold $C_n$; entries marked
  $\sim\!0$ are below $10^{-12}$ in the normalized unbroadened weight.
  }
  \label{tab:dsf_weights}
  \begin{tabular}{c|c|cccc}
  \hline\hline
  $n$ & $C_n$ & $\Gamma$ & $M$ & $K$ & avg \\
  \hline
  1 & $0$ & 0.273 & 0.092 & 0.166 & 0.163 \\
  2 & $\bm{+1}$ & $\sim\!0$ & 0.438 & 0.184 & 0.191 \\
  3 & $0$ & $\sim\!0$ & $\sim\!0$ & 0.105 & 0.093 \\
  4 & $\bm{-2}$ & 0.334 & $\sim\!0$ & 0.191 & 0.145 \\
  5 & $\bm{+1}$ & 0.267 & 0.179 & 0.140 & 0.139 \\
  6 & $0$ & $\sim\!0$ & $\sim\!0$ & 0.072 & 0.090 \\
  7 & $0$ & 0.126 & 0.290 & 0.143 & 0.179 \\
  \hline\hline
  \end{tabular}
\end{table}

\section{Discussion and conclusion}
\label{sec:discussion}

We have used field selection to construct a controlled LSWT reference state in
a frustrated region of the triangular-lattice \KGGp{} model.  At the
representative $\Gamma$-dominant point studied here, the zero-field states do
not converge to a robust compact magnetic cell and are therefore used
only as indicators of finite-size commensuration effects.  An in-plane field along
the crystallographic $a$ axis instead selects a compact noncoplanar seven-site
state over a finite field window, providing the LSWT reference state for the
topological calculation.

Within this seven-site state, bands 5 and 6 form a Dirac-like contact at
$H_c\simeq0.258$ at two generic momenta and exchange Berry curvature across
the transfer, changing the Chern vector from $(0,1,0,-2,1,0,0)$ to
$(0,1,0,-2,-1,2,0)$.  Edge-localized strip modes provide a boundary check of
the bulk Chern sums.  The same transfer appears in the transport response
through Bose-weighted Berry curvature: large band-resolved contributions from
the Chern-active bands nearly cancel, and the reorganization of bands 5 and 6
drives the total $\kxy$ through a near-zero crossing at elevated temperature.
The DSF gives the complementary spectral assignment, showing that the
Chern-active branches carry about half of the coherent spectral weight
averaged along the path, while the Chern transfer itself is more visible in
topology and transport than
in a large redistribution of spectral intensity.

These results are a controlled semiclassical LSWT analysis of the
dimensionless nearest-neighbor \KGGp{} Hamiltonian.  A quantitative
material-specific application would require exchange parameters, $g$ tensor,
interlayer coupling, further-neighbor exchanges, and linewidth effects.
Quantum corrections at $S=1/2$ and the neighboring competing 20- and 36-site
states lie beyond the present leading large-$S$ treatment;
nonperturbative calculations would be needed to test the persistence of the
seven-site stability window and its magnon topology in the deep quantum limit.

\section*{Data Availability}

The data that support the findings of this article are not publicly available.
The data are available from the authors upon reasonable request.

\begin{acknowledgments}
This work was supported by the National Key Research and Development Program
of China (Grant No. 2024YFA1408104), the National Natural Science Foundation
of China (Grant Nos. 12374137, 12434005, and 11804291), and the Natural
Science Foundation of Jiangsu Province (Grant No. BK20241929).
\end{acknowledgments}

\appendix

\section{Magnetic-cell detection algorithm and reciprocal-vector convention}
\label{app:cell_detection}

For each relaxed spin configuration we scan candidate integer translations
$\mathbf{T}_1=m_1\mathbf{a}_1+n_1\mathbf{a}_2$,
$\mathbf{T}_2=m_2\mathbf{a}_1+n_2\mathbf{a}_2$ up to a search radius and
retain the two independent translations of smallest magnetic unit cell that
reproduce the spin pattern to a tolerance $\epsilon=10^{-5}$ on the spin
components.  The retained cell stores the basis-site positions, spin
directions, and nearest-neighbor connectivity used as input for LSWT.

The triangular primitive vectors are non\-orthog\-onal, so the reciprocal
vectors of the magnetic cell are constructed from the real-space supercell
matrix
\begin{equation}
  A_{\rm mag}=(\mathbf{T}_1,\mathbf{T}_2)
\end{equation}
as
\begin{equation}
  B_{\rm mag}=2\pi(A_{\rm mag}^{-1})^{\mathsf T},
\end{equation}
so that the columns of $B_{\rm mag}$ satisfy
$\mathbf{B}_\mu\cdot\mathbf{T}_\nu=2\pi\delta_{\mu\nu}$.  This convention
is used consistently in the band-structure and Berry-curvature calculations.

\section{Magnetic-cell diagnostics at
\texorpdfstring{$(\theta,\phi)=(0.55\pi,1.91\pi)$}{theta=0.55pi, phi=1.91pi}}
\label{app:zero_field_cells}

The zero-field magnetic-cell search at
$(\theta,\phi)=(0.55\pi,1.91\pi)$ gives the cluster-size dependence summarized
in Table~\ref{tab:zero_field_cells}.

\begin{table}[!ht]
  \centering
  \caption{Cluster-size dependence of the detected magnetic cell at
  $(\theta,\phi)=(0.55\pi,1.91\pi)$ and $H=0$.}
  \label{tab:zero_field_cells}
  \begin{tabular}{c|c|c}
  \hline\hline
  $L$ & detected $N_{\rm mag}$ & interpretation \\
  \hline
  24 & rank-one & no two-dimensional cell \\
  30--48 & none & no period found \\
  54 & 108 & finite-size approximant \\
  60 & 20 & isolated lock-in \\
  66 & 132 & finite-size approximant \\
  72 & 72 & finite-size approximant \\
  \hline\hline
  \end{tabular}
\end{table}

In the table, ``rank-one'' means that one magnetic translation is detected but no
independent second translation is found, whereas ``none'' means that no
magnetic translation is detected within the search window.  The finite cells
detected at larger sizes are listed only as commensurate approximants; because
their sizes do not converge with $L$, they are not used as zero-field
spin-wave vacua.

We also tested the harmonic stability of the detected zero-field approximants
using the same bosonic BdG construction as in the main seven-site calculation.
On a $21\times21$ magnetic-zone mesh the three $L=54$ runs, which detect
108-site cells, all have complex LSWT eigenvalues with
$\max|\mathrm{Im}\,\omega|\simeq4.93\times10^{-2}$ and are therefore not
acceptable harmonic vacua.  The isolated 20-site cells at $L=60$ are positive
with $\omega_{\min}=0.190$, while the 132-site and 72-site approximants remain
positive but soft, with $\omega_{\min}$ between $8.2\times10^{-4}$ and
$1.5\times10^{-3}$ depending on seed and cell.  These diagnostics are the
basis for treating zero field as a commensuration problem rather than as a
platform for assigning magnon Chern numbers.

The same cell-detection convention is used for the finite-size compatibility
check of the field-selected seven-site state.  At $H=0.25$, $H\parallel a$,
the compatible tori $L=14,28,42$ reproduce
$E_7/N=-1.17905$ to roundoff, while incompatible sizes lie higher because they
cannot tile the magnetic unit cell.  The corresponding finite-size plot is
included as Fig.~S1 of the Supplemental Material~\cite{SupplementalMaterial}.

\bibliography{refs}

\end{document}